\documentclass[%
 reprint, 
nofootinbib,
 amsmath,amssymb,
 prl
]{revtex4-2}

\usepackage{graphicx}
\usepackage{dcolumn}
\usepackage{bm}

\newcommand{\OO}{\mathcal{O}}

\newcommand{\ex}[1]{\langle #1\rangle}

\newcommand{\RR}{\mathbb{R}}

\newcommand{\maxprob}[2]{\begin{array}{ll}
	\operatorname{maximize} & #1 \\
	\text { subject to } & #2 \\
\end{array}}
\newcommand{\mk}{M^{(K)}}

\begin{document}

\preprint{APS/123-QED}

\title{A Semidefinite Programming algorithm for the Quantum
Mechanical Bootstrap}

\author{David Berenstein}
 \email{dberens@physics.ucsb.edu}
 \affiliation{Department of Physics, UC Santa Barbara}
\author{George Hulsey}%
 \email{hulsey@physics.ucsb.edu}
\affiliation{Department of Physics, UC Santa Barbara}%




\date{\today}

\begin{abstract}
We present a semidefinite program (SDP) algorithm to find eigenvalues of Schr\"{o}dinger operators  within the bootstrap approach to quantum mechanics. 
The bootstrap approach involves two ingredients: a nonlinear set of constraints on the variables (expectation values of operators in an energy eigenstate), plus positivity constraints (unitarity)
that need to be satisfied.
By fixing the energy we linearize all the constraints and show that the feasability problem can be presented as an optimization problem for the variables that are not fixed by the constraints and one additional slack variable that measures the failure of positivity. To illustrate the method  we are able to obtain high-precision, sharp bounds on eigenenergies for arbitrary confining polynomial potentials in 1-D.

\end{abstract}

\maketitle


Solving for the spectrum of Hamiltonians is a very important scientific problem with applications to the study of molecules (quantum chemistry), atomic physics, solid state physics, etc.
Certain applications also require very high precision in the spectrum if one is to understand theoretical aspects of non-perturbative information, like those that  appear when studying resurgent series \cite{Dunne:2014bca}. 
Novel methods that  compute the spectrum of Hamiltonians to high precision are  very useful in these applications.

Recently, the numerical bootstrap has enjoyed renewed attention in its application to quantum mechanical systems, starting with  \cite{Han:2020bkb}. In previous work, we demonstrated the efficiency of the numerical bootstrap in finding rigorous, precise bounds on the energies of eigenstates in one dimensional Schr\"{o}dinger problems \cite{Berenstein:2021dyf,Berenstein:2021loy,Berenstein:2022ygg}. The same setup for other 1-d problems has been studied in \cite{Bhattacharya:2021btd,Aikawa:2021eai,Tchoumakov:2021mnh,Hu:2022keu,Du:2021hfw,Khan:2022uyz,Blacker:2022szo}.
The algorithmic approach (following the ideas of \cite{Lin:2020mme})  performs a search of possible solutions to the truncated bootstrap problem and gives a `Yes/No' answer to their validity.
If a solution survives, one can increase the size of the truncation and keep searching more finely in the set of possible solutions. This search is in a space of many variables which can grow as the size of the truncated problem increases. This type of search is impractical except on search spaces of low dimension, $d_{search}\leq 3$.

In this letter we describe and implement a semi-definite programming algorithm to numerically find an arbitrary subset of the spectrum of a Hamiltonian which overcomes the problem of searching in a high dimensional search space.
We implement it for problems in 1D with a polynomial potential. At each fixed value of the energy $E = \langle H\rangle$, the algorithm is a linear semidefinite program which may be solved polynomially in the size (depth) $K$ of the constraint matrices. One then scans only over $E$.

\section{\label{sec:sdps}The bootstrap as an SDP}
The quantum mechanical bootstrap works as follows. We start with a Hamiltonian $H$ with a point spectrum. For simplicity we will assume that the potential is polynomial and that the system is one dimensional so that
\begin{equation}
     H= p^2 +V(x)
\end{equation}
From this, we assume that we have an eigenstate of the Hamiltonian with energy $E$. The question of the bootstrap is to decide if $E$ is an allowed eigenvalue of the Hamiltonian or not. To do so
we generate a recursion for the positional moments $\ex{x^n}$ from the two constraints
\begin{equation}\label{eq:cnstrs}
    \ex{[H,\OO]} = 0;\quad \ex{H\OO} = \ex{H}\ex{\OO} = E\ex{\OO}
\end{equation}
which assume that the state is an eigenstate of energy $E$.
Given a collection of such moments, any positive function $|\sum \alpha_i x^i|^2$ will have a positive expectation value. This is a unitary constraint: it states that the probability density associated to the state with energy $E$ is non-negative. The constraint is a quadratic function of the $\alpha_i$ and gives rise to a positive definite matrix $M \succeq 0$ computed from the positional moments. A solution is an allowed state if $E$ and the moments satisfy all the constraints and the positive condition on $M$.   

More generally, beyond 1D, for any operator ${\cal O}$, the expectation value of the positive operator
$\langle {\cal O}^\dagger{\cal O}\rangle\geq 0$ must be non-negative. This gives rise to a positive definite matrix when we pick $\cal O$ from the span of a subset of basis operators.  In 1-$D$ problems the expectation values appearing in $M$ defined above are generally strong enough to determine uniquely the solutions.

If $E$ is a variable determined from the other ones, these latter constraints are nonlinear in the moments $x_n \equiv \ex{x^n}$. One may choose to omit the nonlinear constraints and be left with a linear problem; this is the route in \cite{PhysRevLett.108.200404,Lawrence:2021msm}, where one minimizes the value of the energy given some positivity constraints. The tradeoff is that one is only able to solve for the ground state in the absence of the nonlinear constraints.
An alternative approach to linearization to the one we take here is to apply a convex relaxation of the non-linear constraints in \eqref{eq:cnstrs}. Such a method has been applied in the study of the large $N$ bootstrap \cite{kazakovYM,kazakovMM} to relax non-linearities in the Yang-Mills (or matrix model) loop equations that arise from factorization.

{\bf Fixed-energy recursion.}
A simple way to linearize the problem is to fix the value of energy $E$ and test if $E$ is an allowed value. At each fixed value of the energy the recursion is linear in the $x_n$. Consider an arbitrary potential of even degree $d$:
\begin{gather*}
    V(x) = \sum_{n=1}^d a_nx^n
\end{gather*}
The recursion relates moments $x_n$ with $n \geq d$ to lower moments. For $m \geq 0$ it may be written
\begin{multline}
    \label{eq:recursion}
    x_{d+m} = \frac{1}{2a_d(d+2m + 2)} \left[ \frac{}{}4(m+1)Ex_m \right. 
    \\+ \left. m(m^2-1)x_{m-2}
    -2\sum_{n=1}^{d-1}(n+2m + 2)a_nx_{n+m}\right]
\end{multline}
Generically, initializing the recursion requires the energy as well as the first $d-1$ moments, with $x_0 = 1$. 

The basic object of interest in the bootstrap is the $K \times K$ Hankel matrix with elements $M^{(K)}_{ij} = x_{i+j}$, where $0 \leq i,j \leq K-1$. The unitarity constraint is that $M \succeq 0$; $M$ defines a covariance matrix which must be positive semidefinite.

Before applying the recursion, we may write $\mk$ as a linear function of the first $2K-2$ moments: 
\begin{gather*}
    \mk = \sum_{m=0}^{2K-2}x_m\mathcal{B}_m \succeq 0
\end{gather*}
where the matrices $\mathcal{B}_m$ define the Hankel structure:
\begin{equation*}
    \mathcal{B}_m = \begin{cases}
    1 \text{ if } i+j = m,\ 0 \leq i,j \leq K-1\\
    0 \text{ otherwise}
    \end{cases}
\end{equation*}
The recursion \eqref{eq:recursion} relates the $x_m$ with $m \geq d$ to those with $m < d$; it thus defines another set of symmetric $K \times K$ matrices $F_n(E)$ by the equality
\begin{gather*}
    \sum_{m=0}^{2K-2}x_m\mathcal{B}_m = \mk = \sum_{n=0}^{d-1}x_nF_n(E)
\end{gather*}
As $K \to \infty$, the Hankel matrix $\mk$ defined above will be positive definite only for $E$ in the spectrum of $H$: this has been shown in examples and is expected to be true.  No complete proof exists. For the purposes of this paper we will take that statement at face value. Finite $K$ is a truncation of an infinite set of constraints. We expect the Hankel matrix to be positive definite in some disjoint set $S_{K} \subset \RR$ which strictly contains the spectrum of $H$. Moreover,
the $x_n$ are uniquely determined by $E$.
Numerical experiments \cite{Berenstein:2021loy,Lawrence:2021msm} have shown that the convergence to the eigenvalues (and the moments) is exponentially fast in the size of the truncation.
Furthermore, $S_{K+1} \subset S_K$, etc. This same weak convergence property allows efficient search strategies in a bootstrapping algorithm. 

The main problem in previous explorations of the quantum mechanical bootstrap is that a search is done both in $E$ and in the moments. If there are many moments that are undetermined from the recursion, the search for solutions of the bootstrap equations and constraints is
done in a high dimensional space and becomes very inefficient. Our goal then is to find an optimal value of the moments for fixed energy $E$ rather than doing a blind search. Moreover, if the problem fails to find a solutions of the constraints, we want a numerical measure of how far we are from satisfying the constraints.
Our new proposal addresses these issues, so that in end then one is left only with a scan over energies $E$. 

{\bf Optimization.} How do we test if a symmetric matrix $\mk$ is positive? If the matrix is Hermitian, then the condition of being positive (definite) is equivalent to the minimal eigenvalue of $\mk$ being positive. 
We test positive definiteness by considering the minimal eigenvalue of $\mk$ as a function of the primal variables $x_i$. Define an optimization problem
\begin{equation}
    \operatorname{maximize }\  \lambda_{\text{min}}(
    \mk(x_i),E)
    \label{eq:maxprob}
\end{equation}
If the optimal value is negative, the energy value $E$ can be safely excluded from the set $S_K$. The goal is to solve this optimization problem for a range of energies and to thereby determine the set $S_K$. The algorithm proceeds by searching this set at depth $K + 1$, and iteratively converges to the spectrum (or a subset thereof).

The problem \eqref{eq:maxprob} defines an objective function which is highly nonlinear in the primal $x_i$. However, the problem of eigenvalue extremization is well-known to have an equivalent formulation as an SDP with linear objective \cite{boyd}. First, introduce a slack variable $t$ and write
\begin{equation*}
    \maxprob{t}{\lambda_{\text{min}}(M(x_i))\geq t}
\end{equation*}
which is equivalent to \eqref{eq:maxprob}. It is convenient to introduce the matrix 
$M-t I$. If the minimal eigenvalue of $M-t I$ is positive, the matrix in question must be positive 
definite $M-t I \succeq 0$.
This allows us to write a problem equivalent to \eqref{eq:maxprob} in SDP form:
\begin{equation}
    \maxprob{t}{M(x_i) - tI \succeq 0}\label{eq:sdp}
\end{equation}
This is an SDP in linear matrix inequality (LMI) form  with primal variables $\mathbf{x} = (t,x_1,...,x_{d-1})$ \footnote{In some SDP solvers, the algorithm must be written as a minimization problem. This is done by minimizing $-t$ instead.}.


Notice that even if the energy is not allowed, the optimization problem will find a solution: a sufficiently large negative $t$ will always make it possible to satisfy the positive matrix constraint. We thus obtain for the $K$ we are testing a value $t$ that is negative and an optimal value of the moment variables. 
The maximum $t$, which we label $t_{max}$, is a measure of how close to success we are. As we scan over $E$ (at fixed $K$),  $t_{max}$ will depend continuously on $E$ and it is possible to estimate when it will become positive. It thus serves not only as a diagnostic of failure, but it also gives a way to scan intelligently in $E$.

{\bf Problems on other domains.} 
For problems on the half line, the interval, or a circle, light modifications of the approach are needed. In the case of the circle, one uses periodic functions in the bootstrap (a trigonometric moment problem). The goal in that case is to find the band structure of the potential.

There are two main differences from problems adapted to the real line and the interval: certain terms in the recursion are modified and one has  two or more matrix positivity constraints to contend with.

In \cite{Berenstein:2022ygg}, we showed how solving Schr\"{o}dinger problems on the half line requires adding anomalous terms to the recursion which depend on the boundary conditions $\psi(0),\psi'(0)$. One must include these terms, which generally modify the recursion \eqref{eq:recursion}. The same will be true in the interval, where each boundary will modify the recursion relations depending on the boundary conditions.

On the half line, the other difference is due to the result of Stieltjes on the moment problem for measures on $\RR_+$. Positive semidefiniteness is required for the matrix $M_{ij} = x_{i+j}$ as well as the matrix $M'_{ij} = x_{1 + i + j}$. To account for this, we simply make the replacement
\begin{gather*}
    M(x_i) \mapsto \left[\begin{array}{cc}
       \mk(x_i)  & 0 \\
        0  & M'^{(K)}(x_i)
    \end{array}\right] \succeq 0
\end{gather*}
Positive definiteness of the block matrix above is equivalent to positive definiteness of its block-diagonal components. The rest of the algorithm is unchanged, though the size of the constraint matrices will double as they also reflect the block structure. In the interval $(0,1)$, the polynomial $(1-x)$ is also positive definite and there will be additional 
blocks required for solving the dynamics. 

For problems in higher dimensions, we expect that the constraints are not enough to determine recursively all the moments from a finite search space. We are currently investigating this issue. Conceptually, there is no obstacle to proceed in these higher dimensional setups. The main issue will be on understanding the optimal way to eliminate variables and how different truncation schemes might perform.

\subsection{\label{sec:algo}The algorithm}
With the SDP formulation, the bootstrap algorithm proceeds as follows. Given a potential $V$, take an initial set of energy values $S_0 = \{E_i\} \subset \RR$. For each fixed value of the energy, solve the SDP \eqref{eq:sdp} at some initial depth $K_0$. Energies $E_i$ for which the $t_{max}$ is positive form the set $S_{K_0}$, which serves as the search set at depth $K' > K_0$. Iterating this procedure will result in a set of intervals within $S_0$. These intervals define sharp bounds on the exact spectrum of $H$, in the sense that the bounds are rigorous and can only shrink as $K$ increases.

A persistent issue with the bootstrap is the rapid growth of the matrix elements. The magnitude of the largest matrix entries scales super-exponentially with $K$. For example, in the harmonic oscillator, $\ex{x^n} \sim \Gamma(n/2)$ in eigenstates. As a result, using single or double precision floats results in serious numerical error after $K \sim 10$. Similar issues were encountered in the conformal bootstrap program, which necessitated the use of an arbitrary-precision SDP solver \cite{simons-duffin}. We found the same to be necessary in order to obtain comparably high precision to finite-element methods. 

To numerically solve the problem, we used SDPA-GMP \cite{sdpa}, a primal dual interior point SDP solver built on the GMP (GNU multiple precision) arithmetic library. For a given energy, the $F_n(E)$ were generated in Python and the SDP \eqref{eq:sdp} was fed into SDPA-GMP. The outputs defined a refined search space at the next depth. We worked with $\sim60$ digit (200 significant bits)   precision.

The main benefit of the SDP approach is that we can search a very high dimensional space very efficiently. In our previous work, we were constrained to potentials of degree $\leq4$ due to the brute-force nature of the algorithm. Now, potentials of essentially arbitrary degree can be solved in comparable time. 

\section{\label{sec:results}Results for an example}
\begin{figure}[!h]
    \centering
    \includegraphics[width = 8 cm]{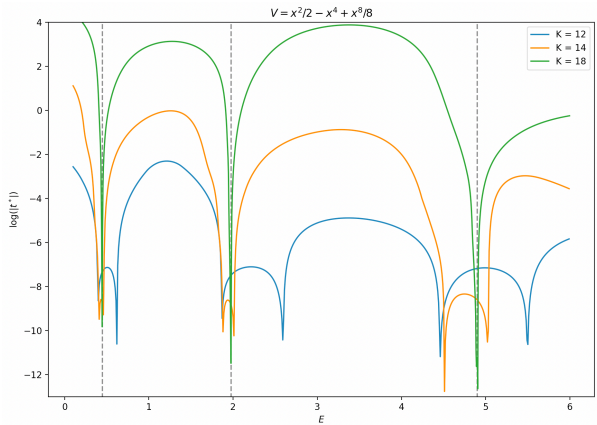}
    \caption{The (log of the) objective function evaluated over a range of energies for the potential \eqref{eq:testpot}. Exact energies (computed in \textit{Mathematica} by FEM) shown as dashed lines. Results shown for $K = 12,14,18$.}
    \label{fig:inaction}
\end{figure}

To show that this method is able to obtain high-precision results for excited states in a search space of large dimension, we considered as a simple example the degree 8 potential
\begin{equation}
\label{eq:testpot}
    V(x) = \frac{1}{2}x^2 - x^4 + \frac{1}{8}x^8
\end{equation}
This has 8 primal variables (including $t$); although since the potential is even, the number effectively reduces to 4 primal variables.
We search over the energy range $[0,15]$ which we know to contain the first five excited states. We started the search at matrices of size $K_0=10$ and terminate at $K=30$. 
At each depth, the algorithm requires us to look for the negative values of the objective function of \eqref{eq:sdp}. We can visualize the convergence by plotting $\log(|t^\star|)$, where $t^\star$ is the optimal value, versus the fixed energy $E$. Inverted `spikes' in this plot show the zero crossings. As the intervals of positive $t$ shrink with increasing $K$, two spikes seem to join around the exact value of the eigenstate energy, as shown in Fig. \ref{fig:inaction}. The structure is always a double spike structure around each allowed value: two spikes can become so close to each other that the plot can no longer distinguish them. 
The numerical estimates for the eigenenergies at $K = 30$ are shown in Table \ref{table:1}. 
\begin{table}[h!]
\centering
\begin{tabular}{c||c|c} 
 \hline
 $n$ & Bootstrap & \textit{Mathematica} FEM \\ [0.5ex] 
 \hline\hline
 1 & 0.446987(6) & 0.44698(8) \\ 
 2 & 1.975515(7)& 1.9755(2)\\
 3 & 4.89758(7) & 4.8975(9) \\
 4 & 9.0514(4) & 9.0514(4) \\
 5 & 14.1008(2) & 14.100(8)\\ [1ex] 
 \hline
\end{tabular}
\caption{Energies for the potential \eqref{eq:testpot} at $K = 30$, compared to the finite-element method (FEM) results. }
\label{table:1}
\end{table}
This level of precision is beyond machine precision in \textit{Mathematica}, though its implementation of a FEM eigensolver works much faster for this class of 1d problems. 

\begin{figure}[!h]
    \centering
    \includegraphics[width = 8cm]{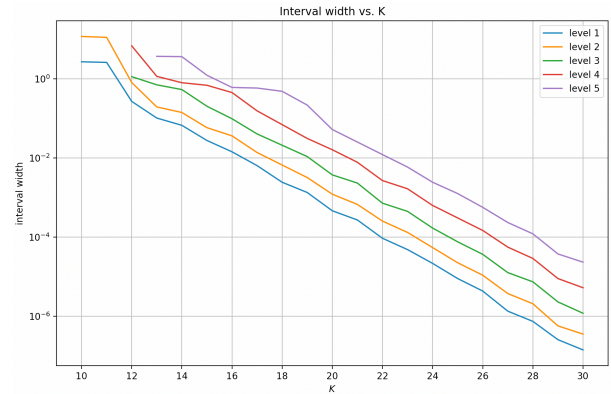}
    \caption{Width of allowed energy intervals vs. $K$, on a logarithmic scale.}
    \label{fig:convergence}
\end{figure}

{\bf Convergence.}
The data from each depth $K$ is a set of valid energy intervals. It has been repeatedly observed that the widths of these intervals decreases exponentially in $K$. We find that result borne out again in Fig. \ref{fig:convergence}.
The convergence is exponential and uniform in slope across energy levels, at least asymptotically in $K$.

In the regime of constant exponential growth of Fig. \ref{fig:convergence}, the approximate slope is $\approx-0.83$. Thus the average width of the allowed intervals decreases like
$
    \Bar{w}(K) \propto e^{-0.83K}
$.
Hence at $K' > K$, the ratio of widths goes like $e^{-0.83(K'-K)}$. Obtaining one more decimal digit of precision requires changing the size of the truncation to $K' = K + \log(10)/0.83 \approx K + 3$. This shows the power of the bootstrap approach: the number of significant digits scales approximately linearly with the depth $K$. 

{\bf Conclusion} In this paper we proposed a method to solve for the energies of 1-dimensional Hamiltonian systems within the bootstrap approach. The method utilizes a semidefinite programming algorithm to find solutions of the (truncated) bootstrap equations. The method solves the problem of ``searches in a large dimension space" by considering the system at fixed energy (the guess) and extremizing over an additional slack variable as well as the other parameters of the original bootstrap equations. What we noticed was that once the energy was factored out, the recursive relations for moments become linear. 
The search space is effectively reduced to one dimension. If the slack variable is positive at the optimal value, the positive definite constraint is satisfied and the energy $E$ is allowed. If the slack variable is negative, in principle one can use a Newton-Raphson method to find the next crossing of zero and thus search effectively in the energy parameter as well. The method is able to obtain high precision data on the eigenvalues and in the example we studied, it is numerically seen that the method converges exponentially fast.

It is clear that our method can be expanded to solving problems in higher dimensions, where the size of the search space might grow with the truncation. Applying these techniques might be useful in the study of many-body problems in quantum chemistry and other areas, with the possibility of not only finding ground state functions of electrons (like in other optimization algorithms \cite{PhysRevLett.93.213001}), but also finding excited states.  

{\em Acknowledgements:} D.B. would like to thank R. Brower,  A. Joseph and J. Yoon for discussions. D.B.  research was supported in part by the International Centre for Theoretical Sciences (ICTS) while participating in the program - ICTS Nonperturbative and Numerical Approaches to Quantum Gravity, String Theory and Holography (code: ICTS/numstrings-2022/9). Research supported in part by the Department of Energy under Award No. DE-SC0019139

\bibliography{refs}

\end{document}